\documentclass[a4paper,11pt]{article}
\usepackage{pos, float}
\usepackage{physics}

\newcommand{\tb}{\theta_{\text{B}}}
\newcommand{\ti}{\theta_{\text{I}}}

\title{Roberge-Weiss transitions at imaginary isospin chemical potential}

\author*[a]{Amine Chabane}
\author[b]{Gergely Endrödi}

\affiliation[a]{Institute for Theoretical Physics, Goethe University, Max-von-Laue-Strasse 1, 60438 Frankfurt}

\affiliation[b]{Faculty of Physics, University of Bielefeld, D-33501 Bielefeld}

\emailAdd{chabane@itp.uni-frankfurt.de}
\emailAdd{endrodi@physik.uni-bielefeld.de}

\abstract{At finite imaginary values of the chemical potential, QCD is free of the sign
problem. Moreover, at high temperatures the partition function exhibits
a new symmetry (the Roberge-Weiss symmetry) connecting phases
with different orientations of the Polyakov loop, and the corresponding
phase transitions between these.
In this contribution we investigate the perturbative one-loop
effective potential for the Polyakov loop in the presence of imaginary
isospin as well as baryon chemical potentials. This leads to a novel phase
diagram, which reveals an interesting insight about the
rich phase structure of the system and the center symmetry breaking.
We check the perturbative results using direct lattice simulations.}

\FullConference{%
 The 38th International Symposium on Lattice Field Theory, LATTICE2021
  26th-30th July, 2021
  Zoom/Gather@Massachusetts Institute of Technology
}

\begin{document}
\maketitle

\section{Introduction}
The study of the phase structure of Quantum Chromodynamics (QCD) at finite temperatures and baryon densities is one of the most important topics in modern physics. A solid understanding of the phase diagram in the $T-\mu$ plane is required for a range of physical phenomena related to the expansion of the early universe, the 
heavy-ion collisions at e.g.\ the Large-Hadron-Collider (LHC) and the properties and structure of neutron stars. 

As of now, we do not have complete knowledge of the QCD phase diagram, neither from a theoretical nor from an experimental perspective. On the theory side, one well established approach 
is perturbation theory which, however, only works at high temperatures, where quarks interact weakly due to the asymptotic freedom property of QCD.
In the strongly interacting regime, the phase diagram can be investigated by means of lattice QCD  simulations. The corresponding Monte Carlo techniques can only be applied for zero (or purely imaginary) chemical potentials. For $\Re\mu\neq0$ the fermion determinant becomes complex and the so-called sign problem appears. There are different methods suggested to circumvent the sign problem, but most of these methods are restricted for to work at $\mu/T \lesssim 1$ (see e.g.\ Refs.~\cite{deForcrand:2009zkb,Aarts:2012yal,Gattringer:2014nxa,Guenther:2020jwe} for reviews). 

\begin{figure}[t]
    \centering
    \includegraphics[scale=0.2]{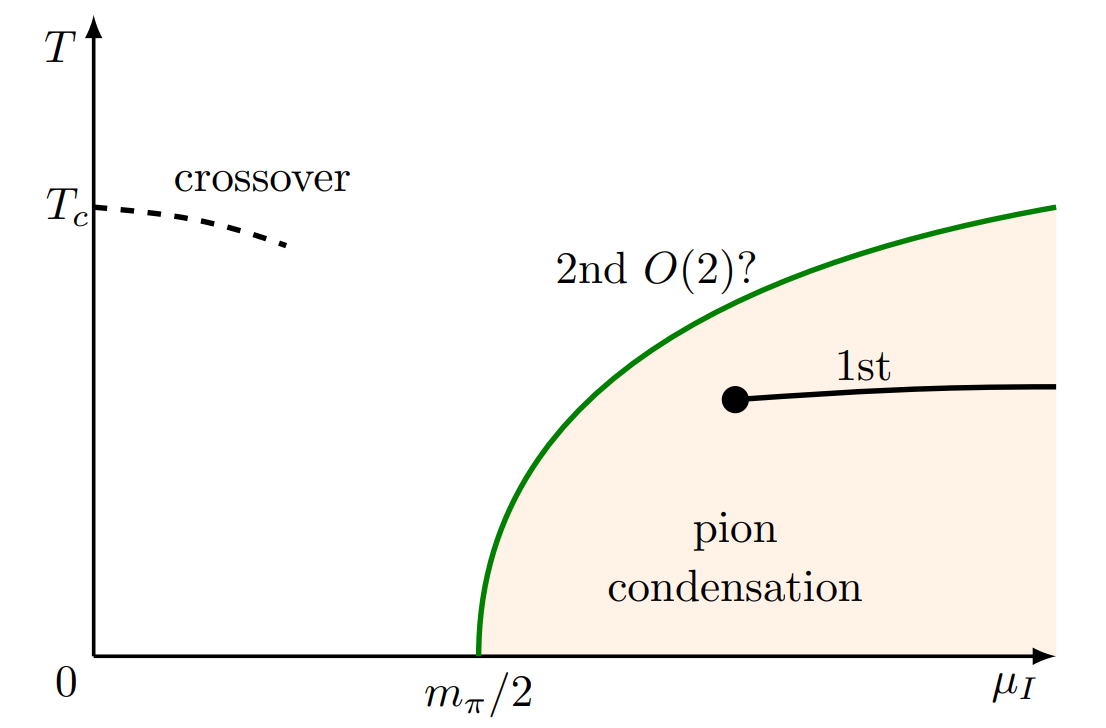}
    \caption{Schematic sketch of the QCD phase diagram in the $\qty(\mu_{\text{I}}, T)$-plane, based on recent lattice simulations~\cite{Brandt:2017zck}.}
    \label{fig:phase_iso}
\end{figure}

Our approach will be to introduce an imaginary chemical potential $\theta \equiv \Im \mu /T$ to avoid the sign problem in a way that, for small $\theta$, we still have the possibility to continue back to real chemical potentials analytically. Furthermore, considering 
the two light quark flavors ($u$ and $d$), we define the baryon $\mu_{\text{B}} = \qty(\mu_{\text{u}} + \mu_{\text{d}})/2$ and the isospin $\mu_{\text{I}} = \qty(\mu_{\text{u}} - \mu_{\text{d}})/2$ chemical potentials and their imaginary counterparts $\tb$ and $\ti$.
While the sign problem only affects nonzero real $\mu_{\text{B}}$, the setup with only nonzero $\mu_{\text{I}}$ is free of the sign problem and standard Monte Carlo simulations can be applied again. The phase diagram in the $\qty(T, \mu_{\text{I}})$ plane has been determined recently~\cite{Brandt:2017zck,Brandt:2018omg} and has provided us with new insights about the structure of QCD. This includes a confirmation of the ``Silver Blaze'' phenomenon~\cite{Cohen:2003kd} at $T=0$ and small $\mu_{\text{I}}$, the 
onset of pion condensation at $\mu_I=m_{\pi}/2$ via a second-order 
phase transition~\cite{Son:2000xc} as well as the emergence of further phases at high $\mu_I$~\cite{Brandt:2019hel} (see Fig. \ref{fig:phase_iso}). 

This contribution is organized as follows: first we briefly review the perturbative treatment of QCD at high temperatures and discuss the behavior of the effective potential in the presence of imaginary chemical potentials. Here we will follow 
the study by 
Roberge and Weiss~\cite{Roberge:1986mm} from 1986. Then we will extend their investigations by introducing imaginary baryon as well as isospin chemical potentials in the effective theory. This will lead to a novel phase diagram, which reveals further details about the phase structure of the system and center symmetry breaking.

\section{Phase structure of $SU(N)$ as a function of the imaginary chemical potential}
In this section we briefly summarize selected properties of the QCD phase structure at finite imaginary chemical potential. 

\subsection{Roberge Weiss Periodicity}
The method of an imaginary chemical potential represents a viable way to bypass the sign-problem and moreover to extend the QCD phase diagram with a new axis. On this axis QCD exhibits a new symmetry which is closely related to the general center symmetry of pure $SU(N)$ gauge theory. Here, $N=3$ denotes the number of colors. It is well known that fermions break the symmetry explicitly, but the effect of a $Z(N)$ transformation on the fermions can be canceled by the shift $\theta \rightarrow \theta + \frac{2\pi k }{N}$ in the imaginary chemical potential. This implies that the partition function $\mathcal{Z}$ is periodic in $\theta$,
\begin{align}\label{RW_periodicity}
	\mathcal{Z}\qty(T, \theta) = \mathcal{Z}\qty(T, \theta + \frac{2\pi k }{N}), 
\end{align}
where k $\in \mathbb{N}$. This phase structure was studied by Roberge and Weiss \cite{Roberge:1986mm} and therefore we talk about Roberge-Weiss (RW) periodicity. In addition, QCD has also special transition points at $\theta = \qty(2k-1)\pi/N$, which are referred to as the RW transitions. The RW transitions are of first order, as also confirmed by lattice calculations, see for example Refs.~\cite{DElia:2009bzj, Bonati:2014kpa}. RW transitions are of first order down to some temperature at which the first order lines terminate into endpoints, where the order of the transition depends on the mass of the quarks \cite{Czaban:2015sas,chris_lat21}.

\subsection{The Polyakov loop}
The Polyakov loop is a gauge invariant quantity defined as
\begin{align}\label{polyakov_loop}
    L(\va{x}) = \frac{1}{N} \Tr \mathcal{P}   \exp\qty{ig\int^{\beta}_{0}\mathcal{A}_{0}(\va{x}, t) \text{d}t},
\end{align}
where $\mathcal{P}$ is the time-ordering operator, $\beta=1/T$ the inverse temperature, $\mathcal{A}_0 = A^{a}_{0} \lambda_a$ and $\lambda^a$ are the generators of $SU(N)$. The Polyakov loop is the trace of a Wilson line winding around the lattice in the imaginary time direction. It is the order parameter for center symmetry. For a constant (background) gauge field $\mathcal{A}_0$, the expectation value of the Polyakov loop will look like
\begin{align}
    \expval{L} = \Tr e^{i\beta \phi},
\end{align}
where 
\begin{align}
    \phi = g \mathcal{A}_0 = \mqty(\dmat{\phi_1,\ddots,\phi_{N}}).
    \label{eq:ploop1}
\end{align}
after diagonalizing the gauge field. The unitarity of $L$ requires that $\sum_a \phi_a=0 \; \textmd{mod} \;2\pi T$.

It is also well known that the Polyakov loop is related to the exponential of the free energy $F_q$ of a static quark~\cite{PhysRevD.24.450}, connecting the expectation value of the Polyakov loop $\expval{L}$  to the confinement and deconfinement properties of QCD.

\subsection{Effective Potential}
The one-loop effective potential for the order parameter $\expval{L}$ for $SU(N=2)$ gauge thery was evaluated in Ref.~\cite{PhysRevD.24.475} and generalized to $N=3$ in Ref.~\cite{PhysRevD.25.2667}. The total effective potential for $SU(N)$ is the sum of the gluonic and the fermionic contributions (assuming one massless fermion) and it is given by  
\begin{align} \label{eff_total}
        V_{\text{eff}}\qty( \phi_{1}, \dots, \phi_{N}) =     V^{\text{glu}}_{\text{eff}}\qty( \phi_{1}, \dots, \phi_{N}) + V^{\text{ferm}}_{\text{eff}}\qty( \phi_{1}, \dots, \phi_{N}),
\end{align}
where 
\begin{align}\label{eff_gluon}
   V^{\text{glu}}_{\text{eff}} \qty(\phi_{1}, \dots ,\phi_{N}) &= \frac{\pi T^{4}}{24} \sum^{N}_{j=1}\sum^{N}_{k=1} \qty{ 1-\qty[\qty[\frac{\beta \phi_{j}}{\pi} -\frac{\beta \phi_{k} }{\pi}]_{\mod 2} -1]^{2} }^{2} \\ \label{eff_ferm}
   V^{\text{ferm}}_{\text{eff}} \qty(\phi_{1}, \dots ,\phi_{N}) &= -\frac{\pi T^{4}}{12} \sum^{N}_{j=1} \qty{ 1-\qty[\qty[\frac{\beta \phi_{j}}{\pi} + 1]_{\mod 2} -1]^{2} }^{2}.
\end{align}
After the discussion on the imaginary chemical potential in the previous section, the next crucial step will be to insert the chemical potential in the total effective potential in Eq.~\eqref{eff_total}. 
Since including an imaginary chemical potential $\Im \mu= \theta/\beta$ merely amounts to a shift of $\mathcal{A}_0$ by $\theta/\beta$ times the identity matrix,
we just need to shift all eigenvalues equally: $\phi_i \rightarrow \phi_i + \theta/\beta$. 
The gluonic part is invariant under this shift, because the chemical potential cancels out in the inner bracket of Eq.~\eqref{eff_gluon}. In turn, the fermionic potential depends explicitly on the chemical potential. The result for the total effective potential becomes
\begin{align}
    V_{\text{eff}}\left( \phi_{1}, \dots, \phi_{N}\right) = V^{\text{glu}}_{\text{eff}}\left( \phi_{1}, \dots, \phi_{N}\right) + V^{\text{ferm}}_{\text{eff}}\left( \phi_{1}+\frac{\theta}{\beta}  , \dots, \phi_{N}+\frac{\theta}{\beta}  \right). \label{komplett_eff}
\end{align}  
\section{Results}
\subsection{Behaviour of the effective potential}
\begin{figure}[h]
    \centering
    \includegraphics[scale=0.46]{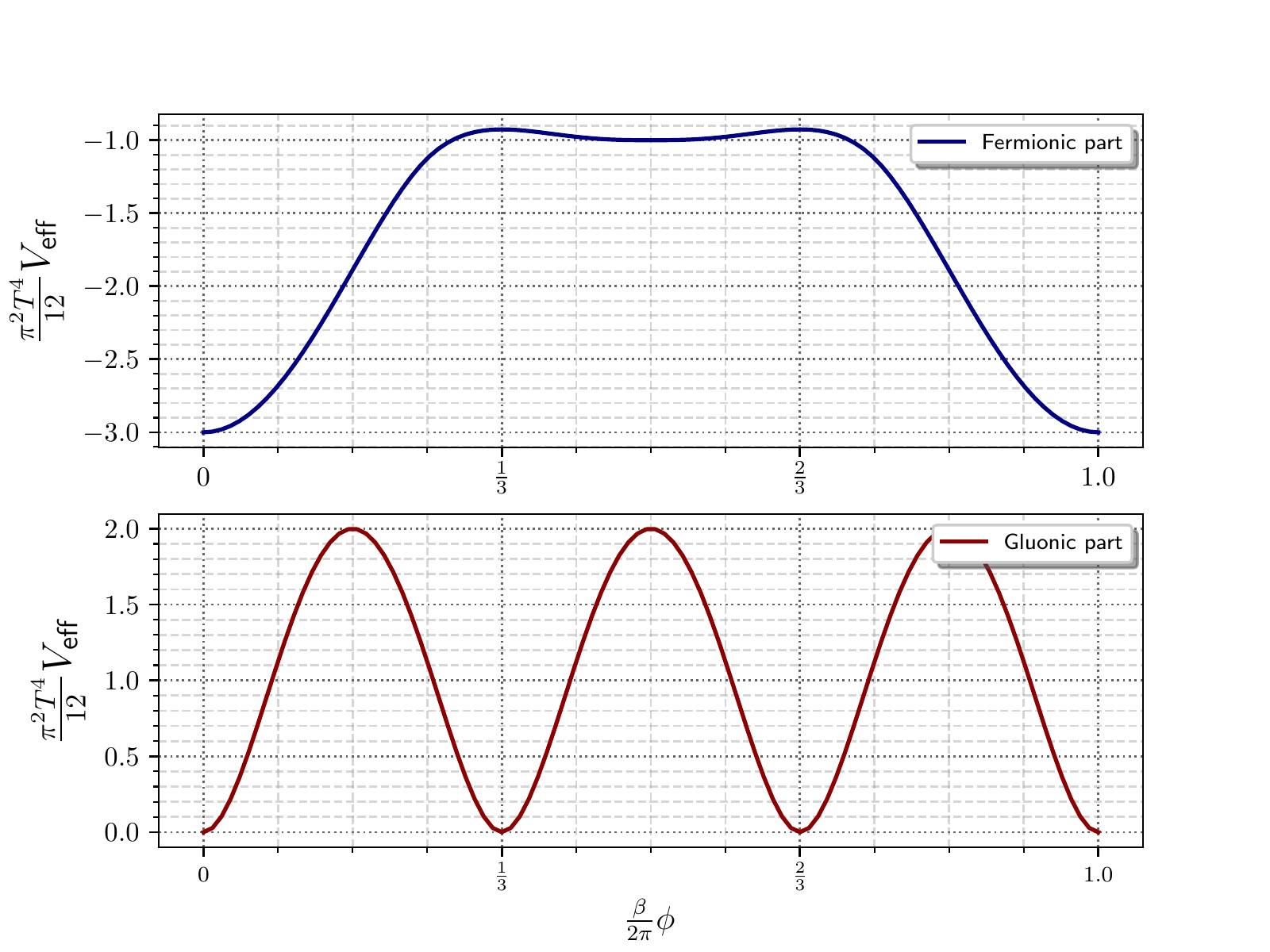}
    \includegraphics[scale=0.46]{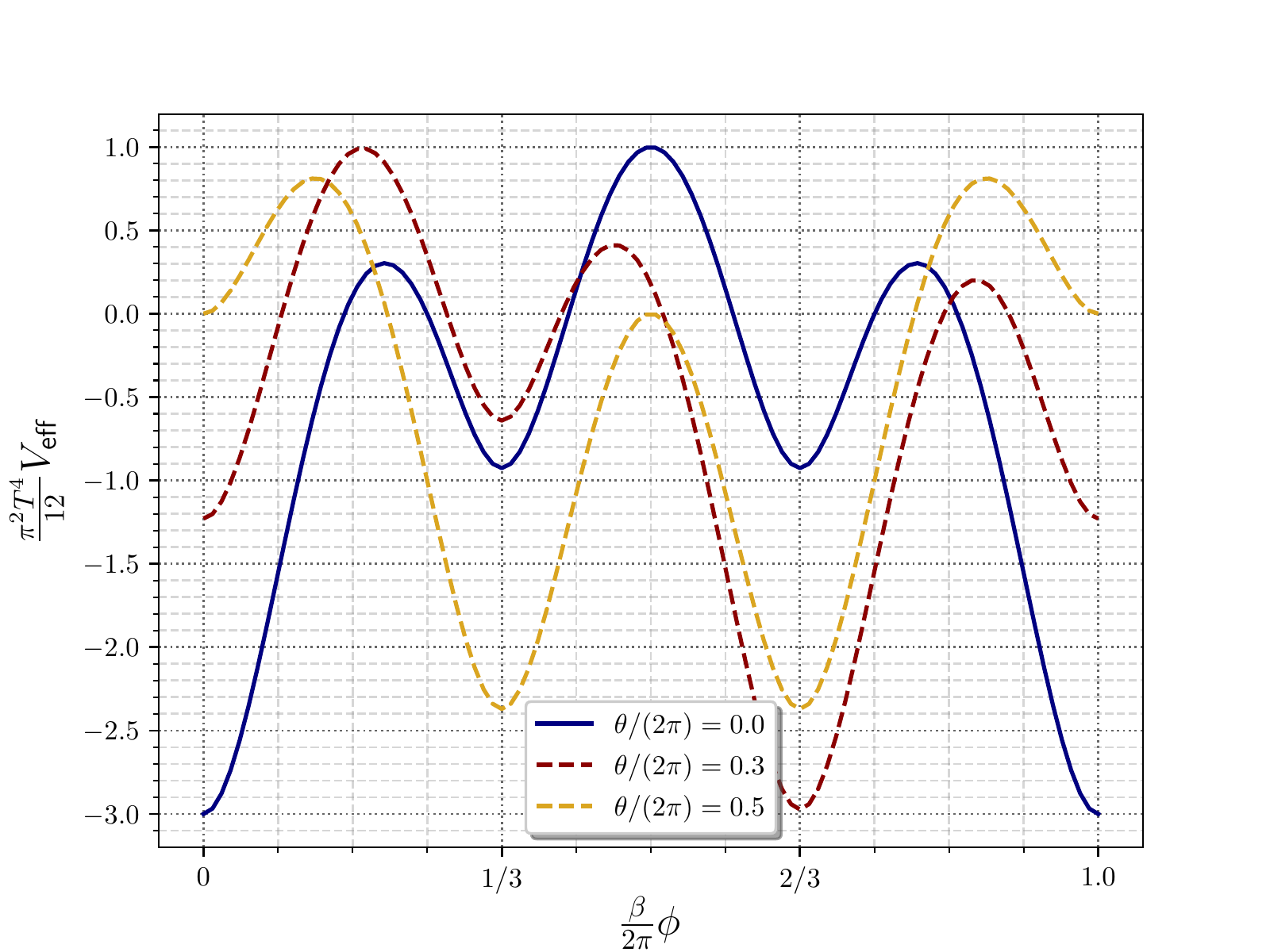}
    \caption{Left: the normalized effective potential $\frac{\pi^2 T^4}{12} V_{\text{eff}}$ for the fermionic and gluonic parts as functions of the fields $\frac{\beta}{2\pi} \phi$. The blue line illustrates the fermionic contribution and the red line the gluonic part of the potential. Right: the total effective potential Eq.~\eqref{eff_total} of the field variables in the same normalization. Here the imaginary chemical potential is now taken into account: $\theta/(2\pi)=0$ (blue), $\theta/(2\pi)=0.3$ (red dashed) and $\theta/(2\pi)=0.5$ (yellow dashed).}
    \label{fig:eff_ferm_gluon}
\end{figure}

Next we analyze the shape and the behavior of the effective potential for fields $\phi_1=\phi_2$. (Note that $\phi_3$ is fixed due to unitarity, see the remark below Eq.~(\ref{eq:ploop1}).) Fig.~\ref{fig:eff_ferm_gluon} shows the fermionic as well as the gluonic part of the effective potential. Two features should be noted here: first, the gluonic part (red line) exhibits the center symmetry, as it has three degenerate minima at $\beta \phi/(2\pi)=[0, 1/3, 2/3]$. In other words, the gluonic effective potential 
is invariant under center transformations.

The lower left plot of Fig.~\ref{fig:eff_ferm_gluon} shows the fermionic contribution (blue line) -- this breaks the symmetry, because there are no degenerate minima anymore and there remains only a global minimum located at $\phi = 0$. This reflects that the fermionic contribution of the effective potential is not symmetric under a non-trivial center transformation. Furthermore, in the right panel of Fig.~\ref{fig:eff_ferm_gluon} we plot the sum of the gluonic and the fermionic contributions, as indicated in Eq.~\eqref{eff_total}. In addition to that, this panel also contains the behavior of the potential in the presence of an imaginary chemical potential. In the figure the solid blue line indicates the potential with zero chemical potential and the remaining dashed lines show the changes due to a nonvanishing chemical potential. We can see that the total effective potential changes in a way that the global minimum moves away from $\beta\phi/(2\pi)=0$, either to $1/3$ or to $2/3$. 

\begin{figure}[h]
    \centering
    \includegraphics[scale=0.3]{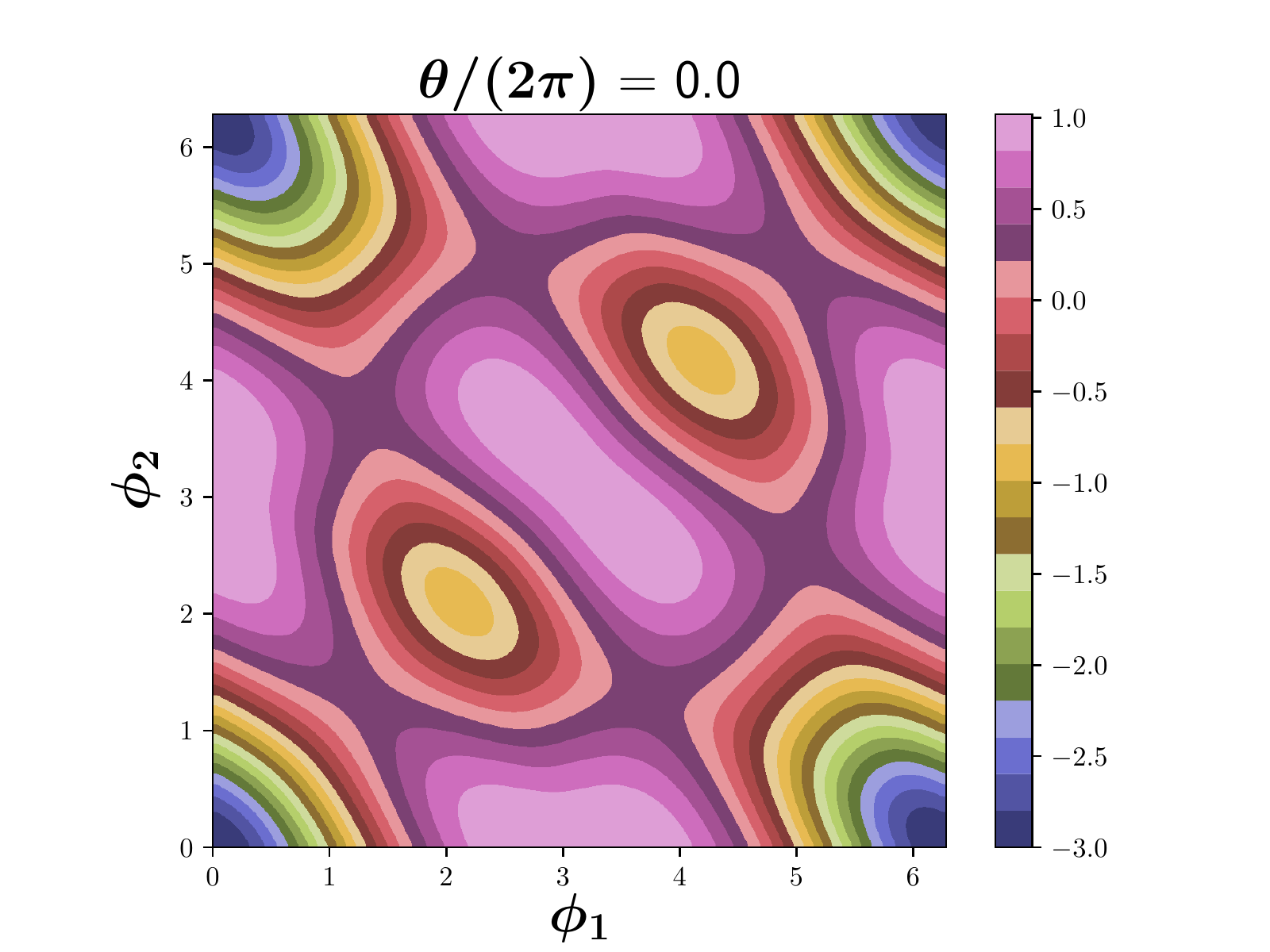}
    \includegraphics[scale=0.3]{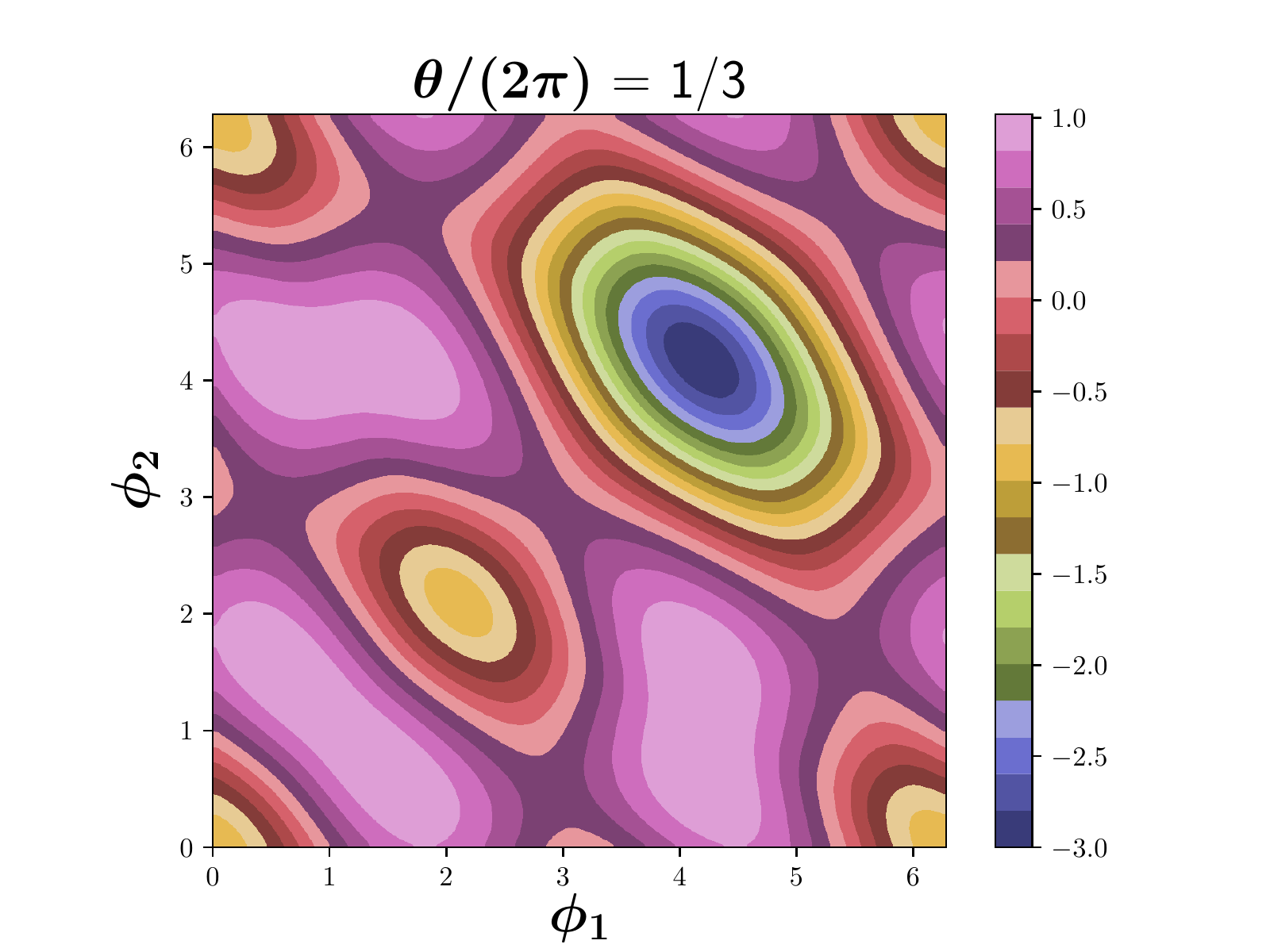}
    \includegraphics[scale=0.3]{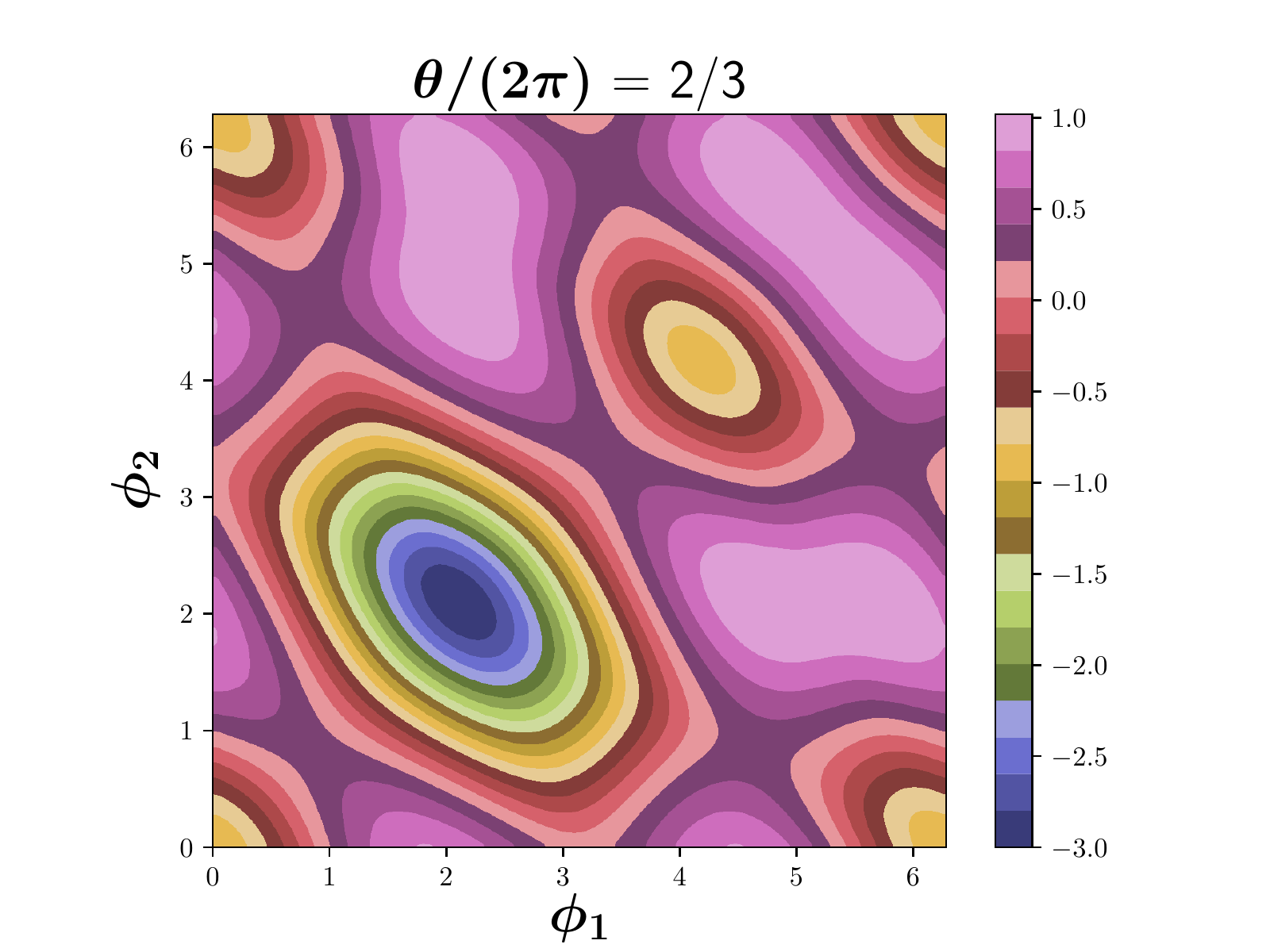}
    \caption{Contour plots of the effective potential $\frac{\pi^2 T^4}{12} V_{\text{eff}}$ in the $\qty(\phi_1 , \phi_2)$-plane for different imaginary chemical potentials $\theta/(2\pi) \in \qty{0, \frac{1}{3}, \frac{2}{3}}$. }
    \label{fig:eff_plane}
\end{figure}

The behaviour with non-vanishing imaginary chemical potential is further studied in  Fig.~\ref{fig:eff_plane}, where the total effective potential is now plotted in the complete  $\qty(\phi_1 ,\phi_2)$-plane for the different imaginary chemical potentials. The first plot (left) shows the effective potential for zero chemical potential, revealing the global minimum at the origin (the other blue areas appear due to periodicity). This is in accordance with Fig.~\ref{fig:eff_ferm_gluon}. Moreover the central and right panel, corresponding to non-vanishing $\theta$, we can observe how the global minimum moves along the diagonal axis. 

\subsection{Adding isospin}
\begin{figure}[h]
    \centering
    \includegraphics[scale=0.46]{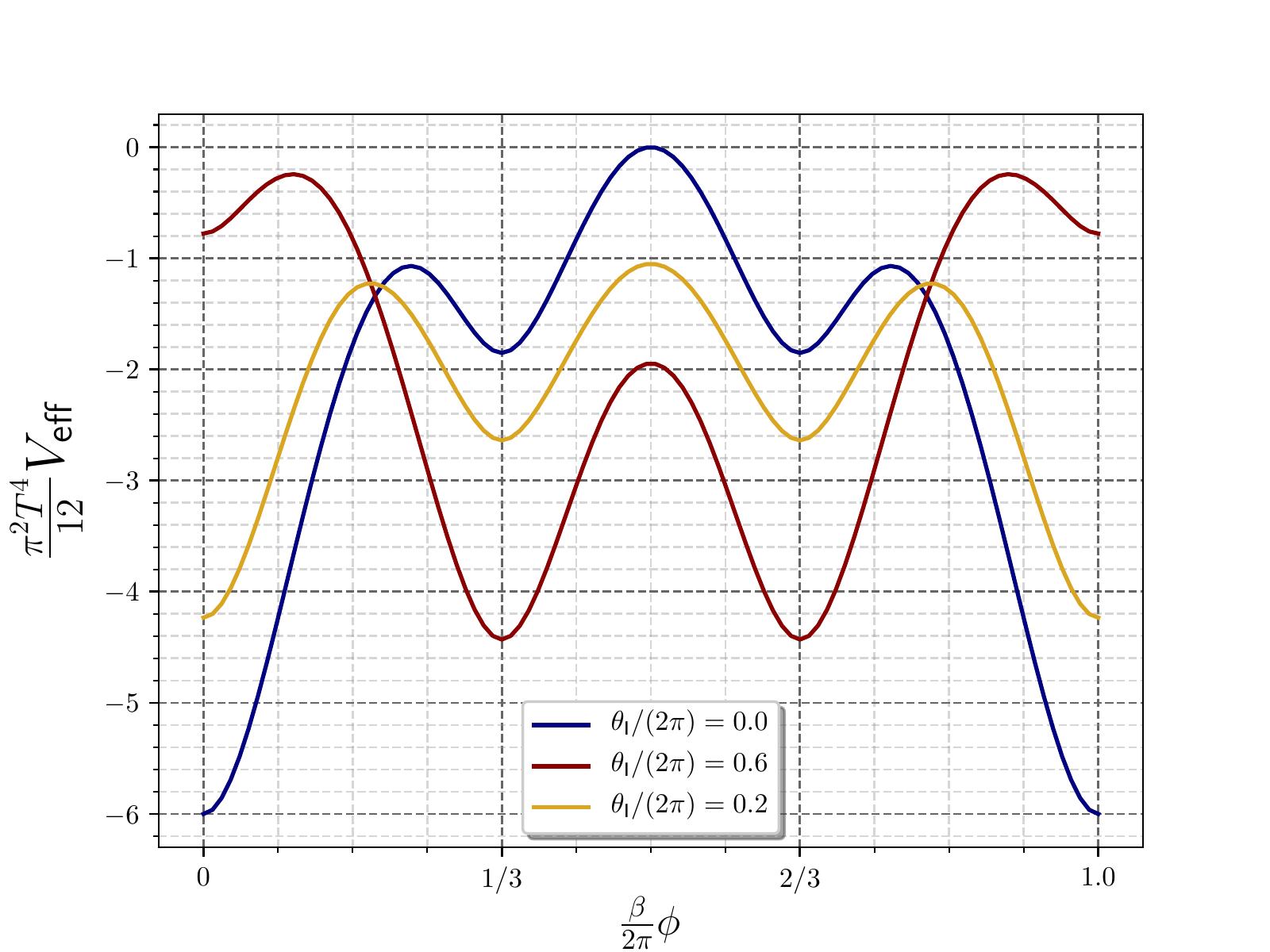}
    \includegraphics[scale=0.46]{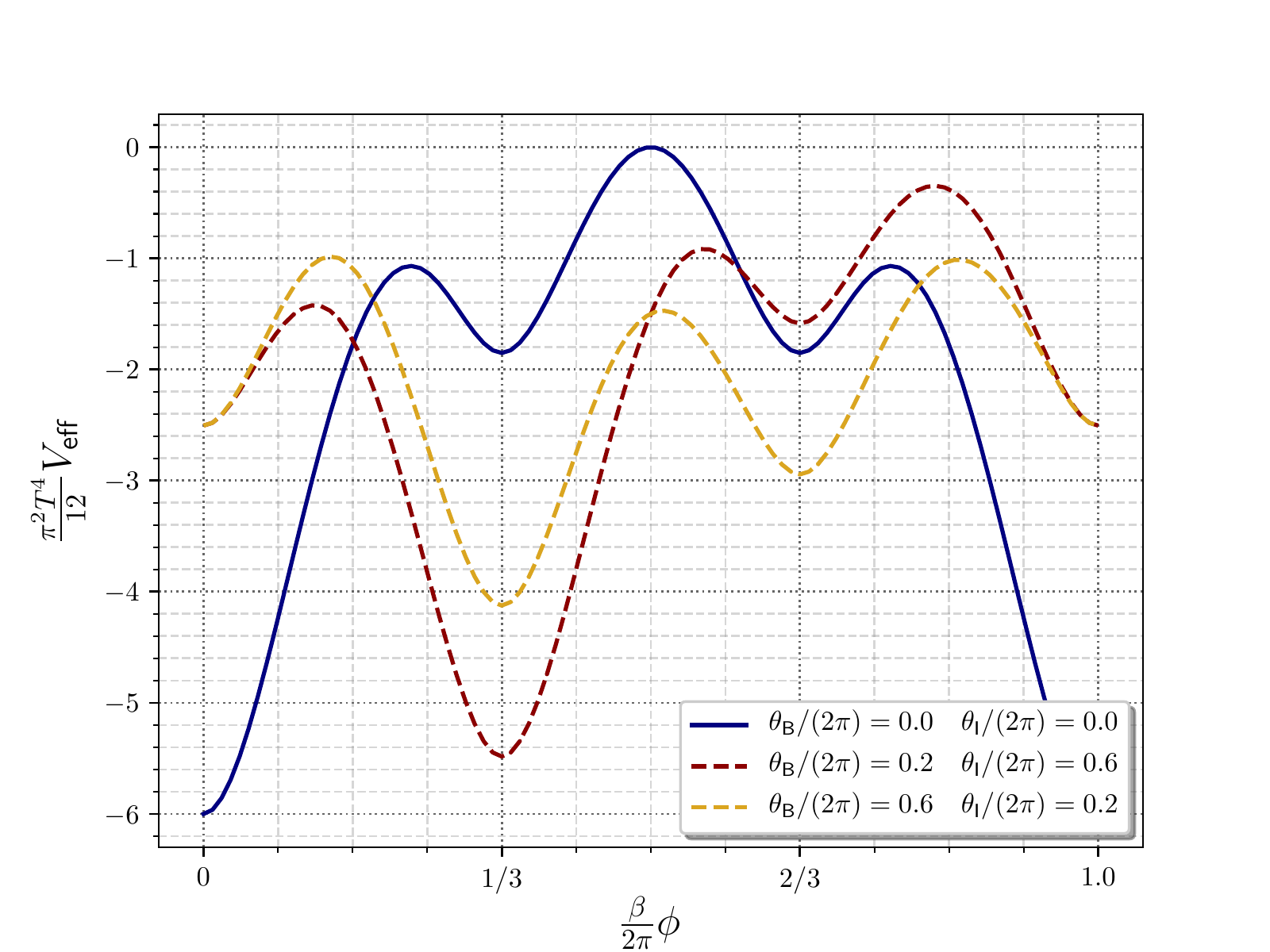}
    \caption{Left: The effective potential $\frac{\pi^2 T^4}{12} V_{\text{eff}}$ as a function of the fields $\frac{\beta}{2\pi} \phi$. Furthermore the isospin chemical potential $\ti$ is now considered  in the effective theory for several values. Right: Again the effective potential $\frac{\pi^2 T^4}{12} V_{\text{eff}}$ as a function of the fields $\frac{\beta}{2\pi} \phi$, where the isospin and the baryon chemical potential are included as a combination. }
    \label{fig:eff_iso}
\end{figure}

After studying the behavior of the total effective potential with and without an imaginary chemical potential, the next step will be to consider two massless quark flavors and distinguish between the baryon $\tb = \qty(\theta_{\text{u}} + \theta_{\text{d}})/2$ and the isospin $\ti = \qty(\theta_{\text{u}} - \theta_{\text{d}})/2$ chemical potentials. The left plot of Fig.~\ref{fig:eff_iso} shows the effective potential as a function of the fields $\phi_1=\phi_2$ for nonzero $\ti$ at $\tb=0$. We can observe that $V_{\rm eff}$ is invariant under the reflection $\phi\to2\pi/\beta-\phi$. After including also the baryon chemical potential (see the right panel of Fig.~\ref{fig:eff_iso}) this symmetry disappears. Now it will be interesting to map out how the global minima change as the baryon and isospin chemical potentials are tuned.

\begin{figure}[h]
    \centering
    \mbox{
    \includegraphics[height=6.3cm]{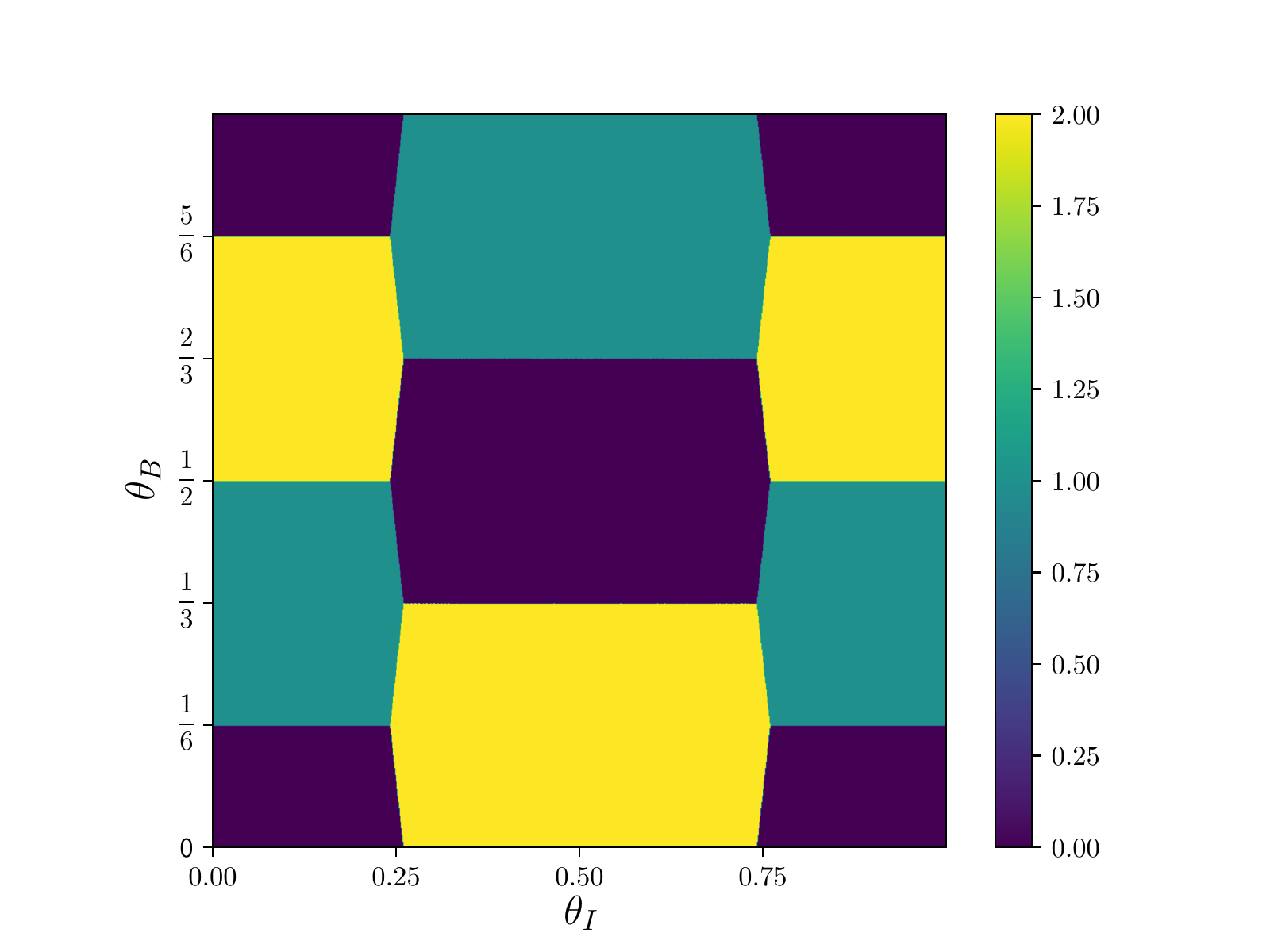}
    \includegraphics[height=6.3cm,trim=60 50 40 60 ,clip]{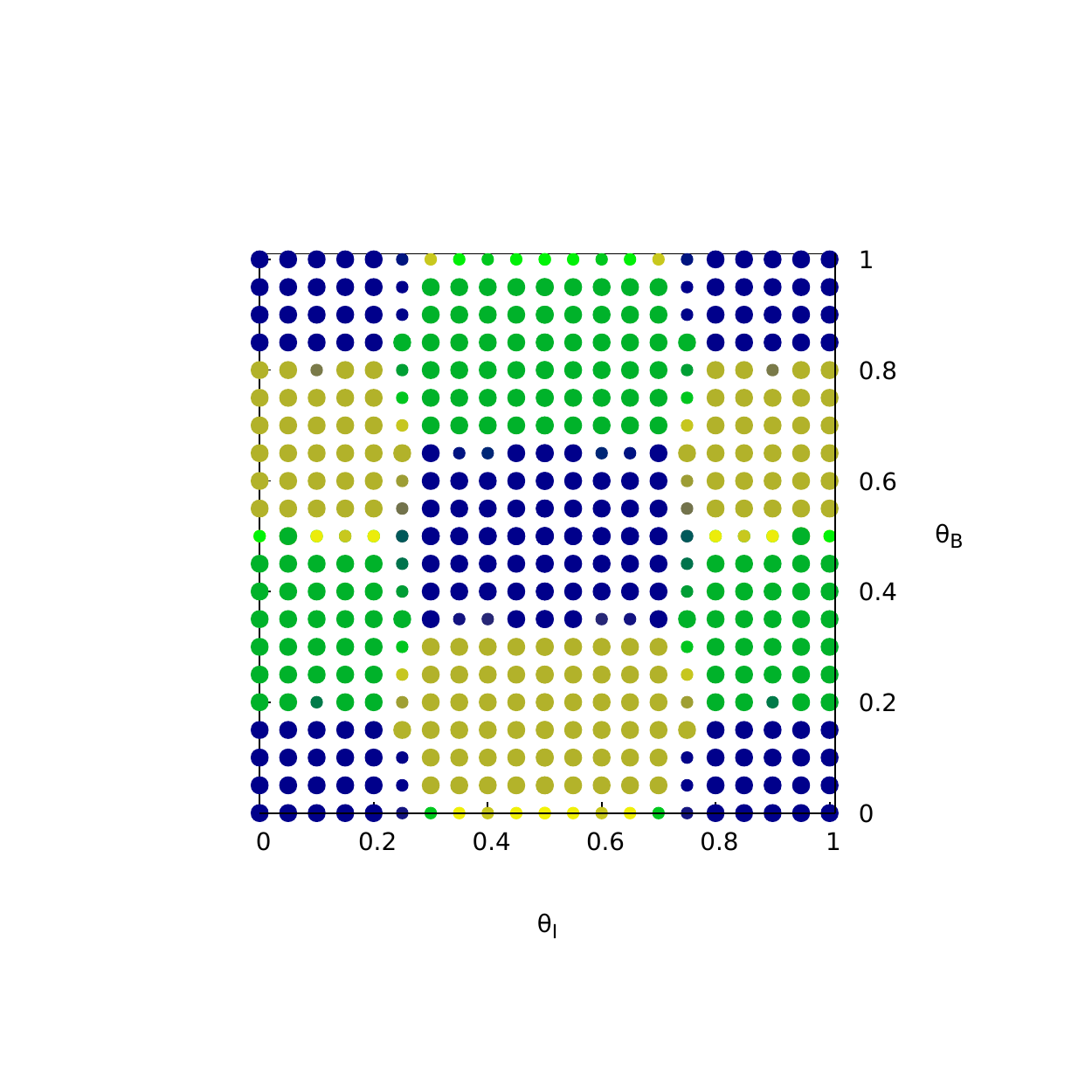}
    }
    \caption{Left: Polyakov loop sectors in the plane of imaginary isospin and baryon chemical potentials, based on the pertubative one-loop effective potential. Right: lattice simulation of the same phase diagram, for details see the text. The three colors represent the three possible phases of the Polyakov loop.}
    \label{fig:carpet}
\end{figure}

The left panel of Fig. \ref{fig:carpet} is our final result for the phase diagram defined from the perturbative one-loop effective potential. It shows the location of the global minimum as a function of the imaginary baryon and isospin chemical potentials. The three different colors correspond to the three sectors of the Polyakov loop ($\beta\phi/(2\pi)=0$, $1/3$ or $2/3$). Along the vertical axis the figure reproduces the 
well-known structure with first-order phase transitions at $\theta_B/(2\pi)=1/6$, $1/2$ and $5/6$. In turn, along the horizontal axis a transition occurs at 
the critical value
$\ti/(2\pi)=\ti^{c}/(2\pi)\approx 0.25602409$ and at $1-\ti^{c}/(2\pi)$, in agreement with Ref.~\cite{Cea:2009ba}. In between these values two phases
coexist (yellow and green colors in the figure). Including both chemical potentials
leads to the repetitive hexagonal pattern.

While at the origin of Fig.~\ref{fig:carpet}, the real Polyakov loop sector 
is preferred and center symmetry is broken, there are special choices for $\tb$ and $\ti$, which effectively restore center symmetry. One special line for example is the horizontal line around the point $\qty(0, 1/6)$, where the blue and green areas coexist. This implies that here the effective potential exhibits a $Z(2)$-type center symmetry, which breaks spontaneously when the system selects one of the two ground states. Another special point is $\qty(\ti^{c}/(2\pi), 1/3)$ where the full $Z(3)$ symmetry of the effective potential is recovered. In this case spontaneous breaking selects one of the three degenerate phases. It is interesting to note that for this choice of chemical potentials the system is center symmetric both with and 
without fermions.

Finally, we check the perturbative results by means of full lattice QCD simulations.
We consider the Wilson gauge action and two flavors of rooted naive staggered quarks with mass $m=0.025$ at an inverse gauge coupling of $6/g^2=5.2$. This setup coincides with that of Ref.~\cite{Endrodi:2014lja}. We simulate $8^{3}\times4$ lattices, corresponding to the deconfined phase at a series of different values 
of $\tb$ and $\ti$, in order to explore the complete phase diagram. A naive hybrid Monte Carlo simulation was found to freeze in incorrect Polyakov loop sectors (i.e., in the one corresponding to a random initial configuration) even in this small volume. In order to solve this problem, we complemented the simulation algorithm with an update step after every twentieth trajectory, offering the system a $Z(3)$ transformation in a random direction, $U_t\to U_t\cdot \exp(\pm i2\pi/3)$ on a random time slice. The action difference is computed exactly via the fermion determinant, using the propagator matrix representation from Ref.~\cite{TOUSSAINT1990248}. A single calculation of the eigenvalues of the propagator matrix is sufficient to get the up and down quark determinant both before and after the $Z(3)$ rotation (this is because the $Z(3)$ rotation coincides with a shift of $\tb$ by $\pm 2\pi/3$). 

The average phase of the Polyakov loop, as measured using the simulations including 
these kind of updates, is shown in the right panel of Fig.~\ref{fig:carpet}. The color coding is the same as in the left panel. For chemical potentials in the interior of the hexagons, the system selects one sector during thermalization and then also remains there for most of the simulation time. These lattice results are shown by the big points. In turn, the small points indicate runs, where frequent jumps between different sectors are observed -- this happens close to the boundaries between sectors, where $Z(2)$ or $Z(3)$ symmetry breaking occurs. We remark that we only calculated the lower left quadrant of the plot was and obtained the remaining parts by symmetry.

Therefore, we can conclude that the lattice data confirms our analytical approach. 
In order to precisely compute the critical isospin chemical potentials, e.g.\ the value of $\ti^c$, simulations on larger volumes are necessary.

\newpage
\section{Summary}

In this contribution we studied the one-loop QCD effective potential in the presence of imaginary isospin and baryon chemical potentials. In particular, we investigated how the global minima of the effective potential change and discussed how center symmetry is (partially) restored for certain choices of the chemical potentials.

Our research provides new information about the role of isospin and barony chemical potentials in the effective theory. Our most important result is the left panel of Fig.~\ref{fig:carpet}. This phase diagram illustrates the Polyakov loop sectors corresponding to the global minimum of the effective potential for all possible combinations of $\qty(\ti, \tb)$. We found that for certain chemical potentials -- lines along the boundaries of hexagons -- $Z(2)$ center symmetry is realized, while at special points -- meeting point of three hexagons -- complete $Z(3)$ symmetry is recovered. In these cases the ground state is chosen spontaneously. Further details are provided in Ref.~\cite{bsc_chabane}.

Furthermore, we checked the perturbative results using direct lattice QCD simulations at high temperature. To this end we needed to include an additional update step in the simulation algorithm, which offers a random center transformation. This was implemented via an exact calculation of the fermion determinant. The obtained data fully confirm the analytical approach (right panel of Fig.~\ref{fig:carpet}), even though we only used a small volume. A precise determination of the critical chemical potentials marking the center sector boundaries will necessitate simulations on larger volumes, likely requiring a stochastic implementation of the center updates.

\acknowledgments
This research was funded by the DFG (Emmy Noether Programme EN 1064/2-1 and the
the Collaborative Research Center CRC-TR 211 ``Strong-interaction
matter under extreme conditions'' -- project number
315477589 - TRR 211). The authors thank Bastian Brandt, Volodymyr Chelnokov and Francesca Cuteri for insightful comments.
 
\bibliographystyle{unsrt}           
\bibliography{biblio.bib}

\end{document}